\documentclass[conference]{IEEEtran}
\IEEEoverridecommandlockouts
\usepackage{cite}
\usepackage{amsmath,amssymb,amsfonts}
\usepackage{algorithmic}
\usepackage{float}
\usepackage{color}
\usepackage{graphicx}
\usepackage{hyperref}
\usepackage{textcomp}
\usepackage{booktabs}
\usepackage{multirow}

\begin{document}

\title{Design Environment of Quantization-Aware \\Edge AI Hardware for Few-Shot Learning\\
}

\author{%
    \IEEEauthorblockN{R. Kanda\IEEEauthorrefmark{1}\IEEEauthorrefmark{2},  N. Onizawa\IEEEauthorrefmark{1}, M. Leonardon\IEEEauthorrefmark{3}, V. Gripon\IEEEauthorrefmark{3} and T. Hanyu\IEEEauthorrefmark{1}}
    \IEEEauthorblockA{\IEEEauthorrefmark{1}Research Institute of Electrical Communication, Tohoku University, Japan}
    \IEEEauthorblockA{\IEEEauthorrefmark{2}Graduate School of Engineering, Tohoku University, Japan}
	\IEEEauthorblockA{\IEEEauthorrefmark{3}IMT Atlantique, Lab-STICC, UMR CNRS 6285, F-29238 Brest, France}
    Email: kanda.ryosuke.q3@dc.tohoku.ac.jp
}

\maketitle

\begin{abstract}
This study aims to ensure consistency in accuracy throughout the entire design flow in the implementation of edge AI hardware for few-shot learning, by implementing fixed-point data processing in the pre-training and evaluation phases.
Specifically, the quantization module, called Brevitas, is applied to implement fixed-point data processing, which allows for arbitrary specification of the bit widths for the integer and fractional parts. 
Two methods of fixed-point data quantization, quantization-aware training (QAT) and post-training quantization (PTQ), are utilized in Brevitas.
With Tensil, which is used in the current design flow, the bit widths of the integer and fractional parts need to be 8 bits each or 16 bits each when implemented in hardware, but performance validation has shown that accuracy comparable to floating-point operations can be maintained even with 6 bits or 5 bits each, 
indicating potential for further reduction in computational resources.
These results clearly contribute to the creation of a versatile design and evaluation environment for edge AI hardware for few-shot learning.
\end{abstract}


\section{Introduction}
As Artificial Intelligence (AI) technology continues to evolve, the computational requirements and energy consumption associated with traditional AI learning have become substantial, necessitating immediate solutions to these challenges. 
Therefore, research on edge AI hardware based on few-shot learning\cite{wang2020generalizing}, which is essentially different from conventional neural network learning\cite{he2015deep}, has attracted attention.

Few-shot learning has recently become a focal point of research in fields such as natural language processing and image recognition. 
This approach enables learning from a limited dataset and applying the knowledge to unfamiliar data, thus potentially providing computers with human-like efficient learning capabilities, especially in situations where data is scarce.

In this study, we employ a Pipeline for Embedded Few-Shot Learning (PEFSL)\cite{PEFSL}, which is an example of few-shot learning implementation for Edge AI. 
The PEFSL framework provides a pipeline for learning on FPGA, hardware synthesis, and the implementation of few-shot learning applications, utilizing the PYNQ-Z1 board and the Tensil framework for executing image recognition AI tasks. 

However, PEFSL currently faces several challenges, one of which includes the discrepancy between pre-training and accuracy evaluation processes, which are performed using floating-point operations, despite the hardware implementation being fixed-point. 
It is unclear whether the accuracy obtained in software will work in hardware implementation. 
In other words, there is no consistent accuracy assurance in the flow, so there is a need to quantize the floating-point sections to fixed-point in order to ensure accuracy consistency.

This paper describes the implementation of fixed-point processing for the pre-training and evaluation phases, which allows for arbitrary specification of the bit widths for the integer and fractional parts.
The implementation was conducted using two approaches: Quantization-Aware Training (QAT) and Post-Training Quantization (PTQ).
In performance, we confirmed that the output of each process was correctly rounded and that the accuracy varied with the specified bit width. 
From these results, this research contributes to the creation of a versatile design and evaluation environment for few-shot learning hardware, considering both accuracy and hardware resources in the implementation.

\section{Framework for Few-Shot Learning}

\subsection{Few-Shot Learning}
\begin{figure}[t]
	\centering
	\includegraphics[width=0.9\columnwidth]{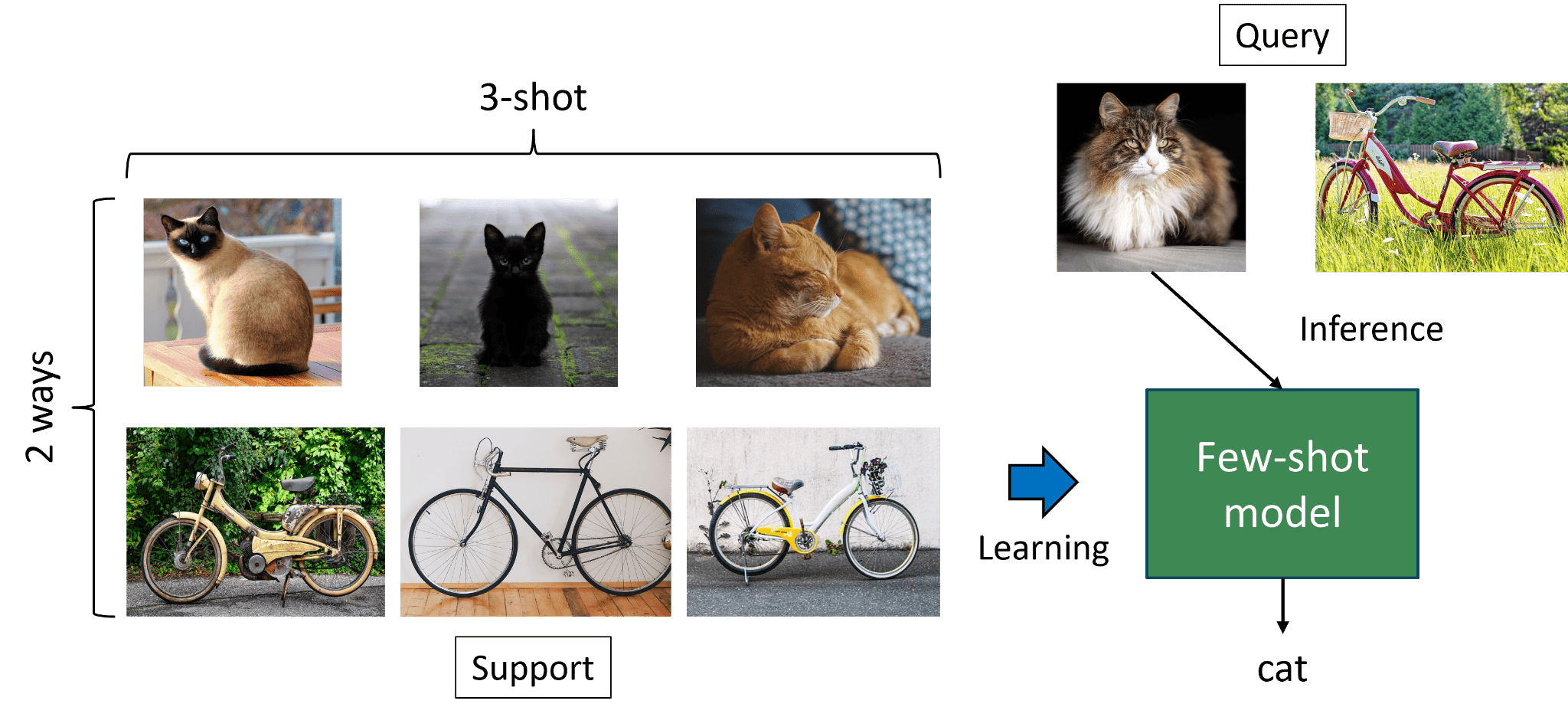}
	\caption{Few-shot learning involves training with a small amount of additional data (support) for each class to be classified, followed by inference using test data (query).}
	\label{about_FSL}
\end{figure}

\begin{figure}[t]
	\centering
	\includegraphics[width=0.9\columnwidth]{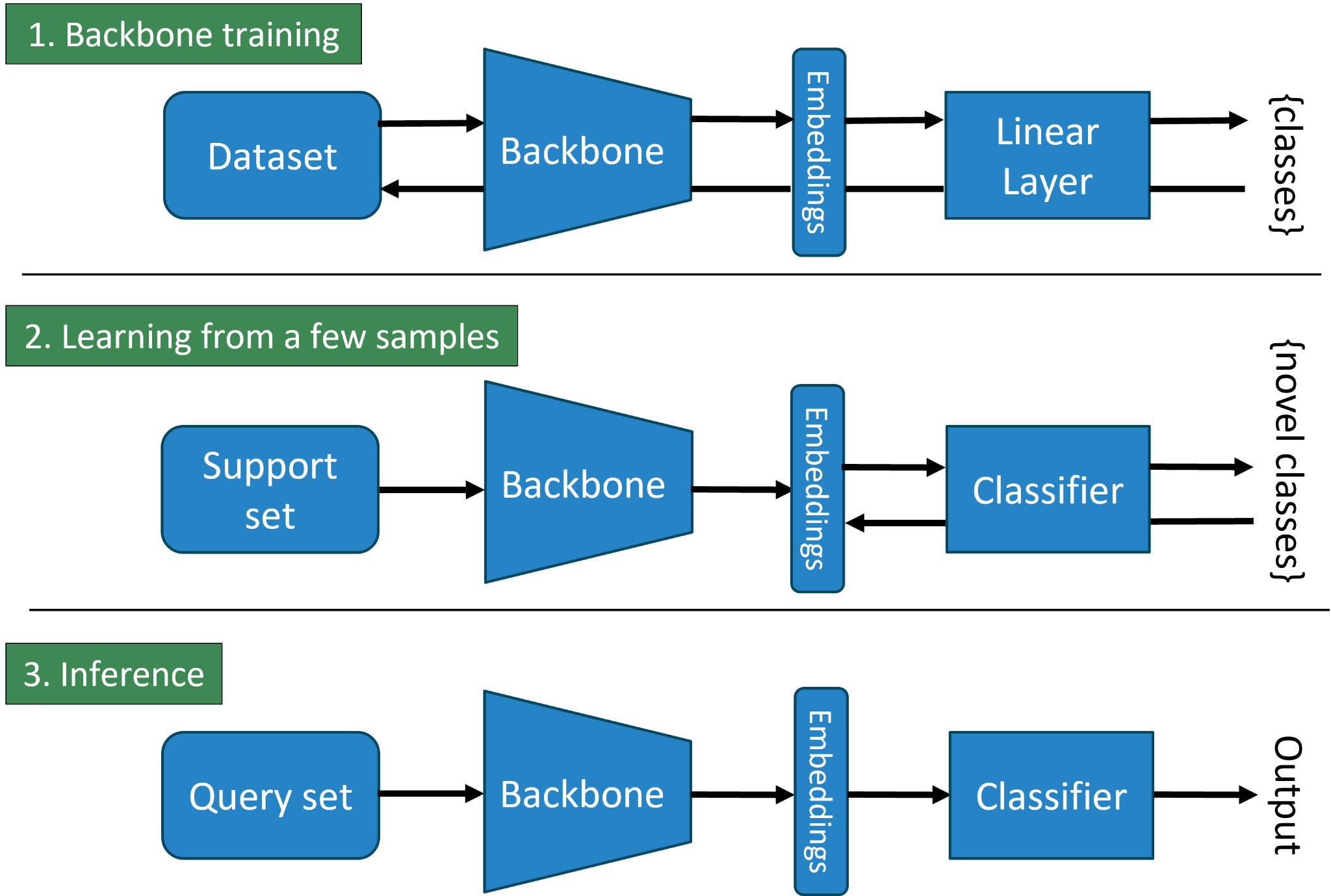}
	\caption{
		Few-shot learning consists of: 
		1. training a backbone to extract features, 
		2. learning features from the trained backbone and a few additional data, and 
		3. executing classification tasks.}
	\label{FSL_flow}
\end{figure}

Few-shot learning is a learning approach aimed at learning from a small number of samples as shown in Fig.\ref{about_FSL}. 
It is  an effective method when the number of data is limited (e.g. medical image diagnosis with a small number of cases).
In few-shot learning, two types of data samples are used: the support set and the query set.
The support set is a small number of labeled samples which the model is trained from, and the query set consists of new samples to evaluate the model's predictive ability.
The learning and inference flow of few-shot learning is as follows: backbone training, learning from a few samples, inference, as shown in Fig.\ref{FSL_flow}.

\subsubsection{Backbone training}
Backbones are trained on large datasets using deep learning models such as Convolutional Neural Networks (CNNs) to create generalized feature extractors.
However, it doesn't include the image classes of the support set to be trained in the next step.

\subsubsection{Learning from a few samples}
The pre-trained backbone is frozen. 
The learning data (support) for few-shot is passed through the backbone to extract features, and training is conducted using a Classifier.

\subsubsection{Inference}
Finally, an image to be classified is first transformed into features using the backbone, then fed to the classifier for generating an output decision.

\subsection{Nearest Class Mean}
There are several approaches to learning environments for few-shot learning, but the learning environment used in this study is an inductive setting.
Inductive setting involves learning from the support set (without learning from the query set) and then making predictions for the query set, as described in the previous section.
In this case, a simple classifier such as Nearest Class Mean (NCM)\cite{wang2019simpleshot}\cite{bendou2022easy} reaches competitive performance, and will be used in our study.

NCM computes the average of the feature vectors of the training data for each class (class mean) and takes these\\ averages as the representative points of the class.
When classifying a new sample, NCM then computes the distance between the sample's feature vector and the representative point of each class, and makes predictions by assigning the sample to the class with the closest distance.

Let us denote $\mathbf{z}$ feature vectors, $S_i(i\in{1, ..., n})$ the set of feature vectors corresponding to the support set for each class, and $Q$ the set of query feature vectors.
Then, predictions are made by computing the barycenters (representative point) of the class from the labeled samples as shown in Equation (1) and taking the difference between each query and the nearest center of the barycenter in Equation (2).

\begin{equation}
	\forall i: \overline{\mathbf{c}_i}=\frac{1}{|S_i|}\sum_{\mathbf{z}\in S_i}\mathbf{z} 
\end{equation}
\begin{equation}
	\forall \mathbf{z}\in Q : C_{ind}(\mathbf{z}, [\overline{\mathbf{c}_1}, ..., \overline{\mathbf{c}_n}]) = \text{arg}\underset{i}{\text{min}}\|\mathbf{z}-\overline{\mathbf{c}_i}\|_2
\end{equation}

\subsection{PEFSL}
As mentioned in the introduction, this study is based on a Pipeline for Embedded Few-Shot Learning (PEFSL), which is an example of edge AI implementation.
PEFSL is designed to enable the execution of few-shot learning on hardware by processing pre-trained data (weight data) obtained through validation and evaluation using Backbone learning and few-shot processing, for hardware implementation using Tensil.

\begin{figure}[t]
	\centering
	\includegraphics[width=0.9\columnwidth]{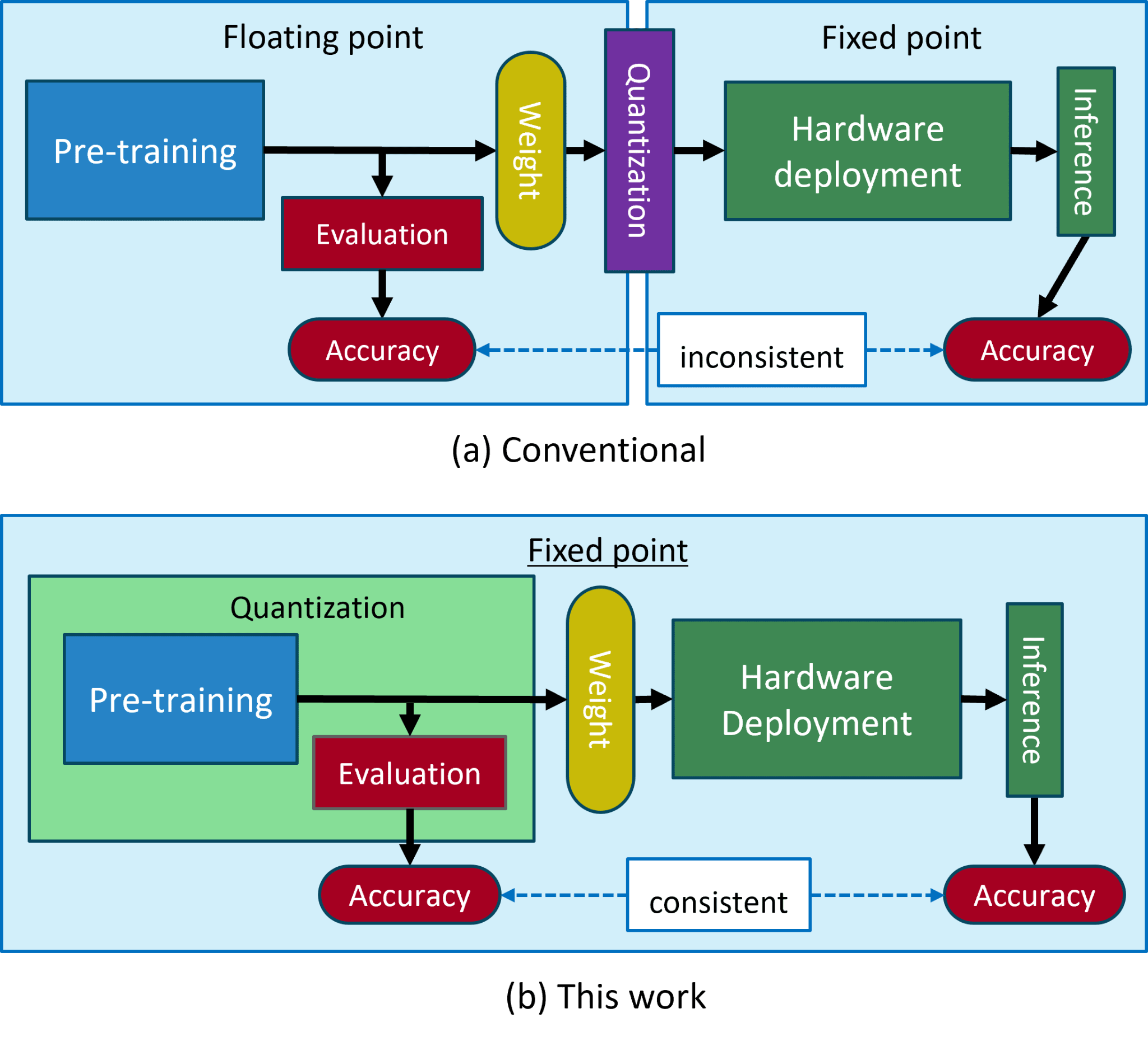}
	\caption{
		In the conventional flow (a), the hardware implementation uses fixed-point processing, while the pre-training phase employs floating-point processing. 
		In the proposed flow (b), by quantizing the entire flow to fixed-point, consistency of accuracy is ensured.}
	\label{flow}
\end{figure}

However, this pipeline has challenges, one of which is that the backbone learning and accuracy evaluation are performed with floating-point processing and have not been quantized to fixed-point representation. 
Fig. \ref{flow} shows a rough flow from pre-training to hardware implementation. 
Although the accuracy evaluation in the pre-training and weight data generation stages is a floating-point process, the resulting weight data is fixed-pointed to a certain bit width by Tensil and implemented in hardware, so the consistency of the accuracy in software and hardware is not guaranteed.
Therefore, this study aims to ensure accuracy throughout the pipeline by quantizing the backbone learning and accuracy evaluation environment to fixed-point representation.

\section{Implementation of Fixed-Point Quantization}
Fixed-point quantization is implemented using Brevitas\cite{brevitas}, \\
which supports two types of quantization techniques: Quantization-Aware Training (QAT) and Post-Training Quantization (PTQ)\cite{gholami2021survey}, with each method  implemented accordingly. 

\begin{figure}[t]
	\centering
	\includegraphics[width=0.9\columnwidth]{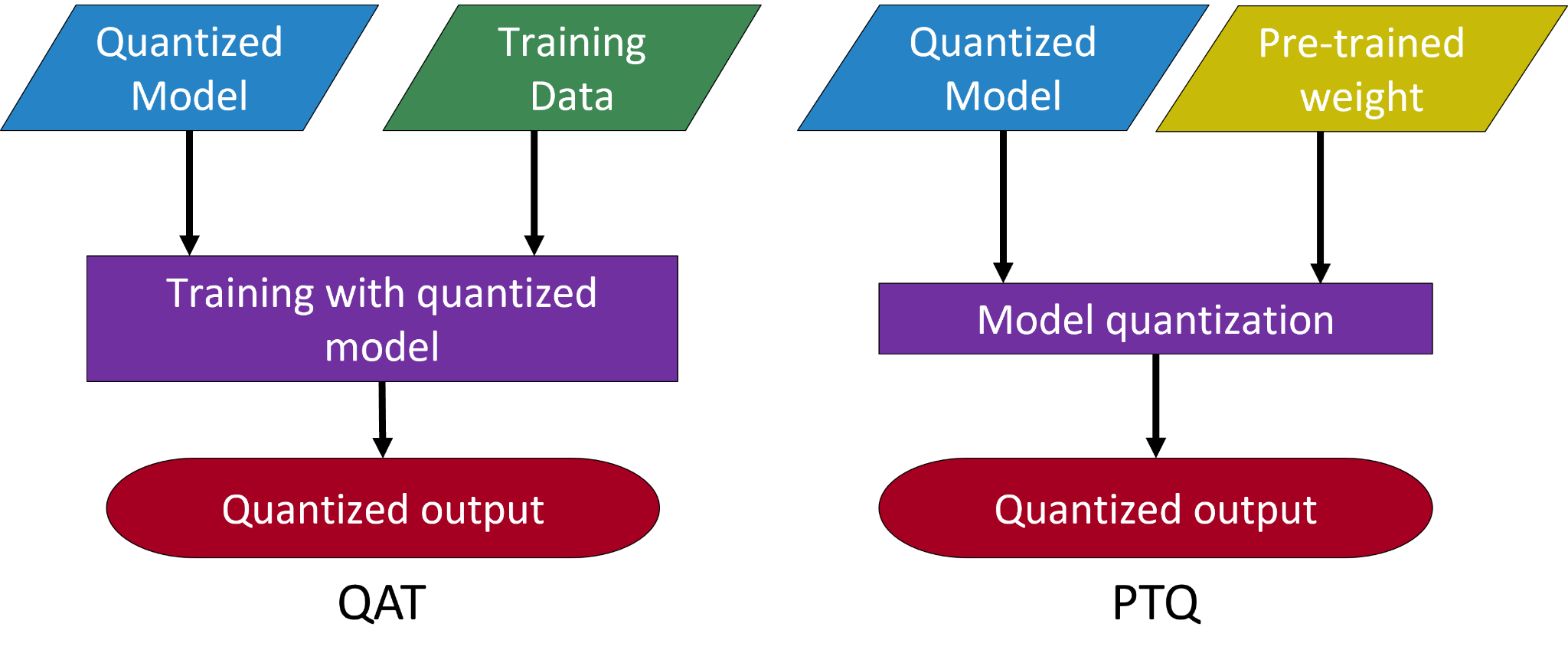}
	\caption{Quantization-Aware Training (left) is a technique that quantizes the model during the training process, allowing the training to account for the effects of quantization, 
	and Post-Training Quantization (right) is a method that applies quantization later to models that have been previously trained with floating-point processing.}
	\label{qat_ptq}
\end{figure}

The difference between QAT and PTQ is shown in Fig. \ref{qat_ptq}.
QAT is a technique that quantizes the model during the training process, allowing the training to account for the effects of quantization. 
By integrating quantization into training, QAT enables the neural network to adapt to lower precision and helps reduce the degradation of model performance when converted to quantized format. 

On the other hand, PTQ is a method that applies quantization later to models that have been previously trained with floating-point processing. 
PTQ requires less processing time and resources than QAT, but may result in a greater loss of model accuracy.
The CNN models implemented in this paper are ResNet12, a commonly used backbone in the field of few-shot Learning, which is based on twelve convolutional layers. 
However, by implementing them in the same way, it is possible to apply this approach to other ResNet models as well.

\subsection{Quantization-Aware Training}\label{QAT_impl}
QAT was implemented by replacing the modules used for convolution and activation layers in the ResNet model provided by PEFSL\cite{bendou2022easy} with Brevitas' quantization modules, and then proceeding with training.

\begin{figure}[t]
	\centering
	\includegraphics[width=\columnwidth]{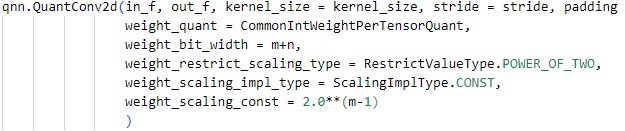}
	\caption{
		Implementation of the convolutional layer. 
		Fixed-point processing is activated by the arguments of the module.}
	\label{quantconv}
\end{figure}

Fig.\ref{quantconv} shows an example of Implementation of the convolutional layer. 
Originally, convolution processes were defined by torch.nn.Conv2d. We replaced them with the quantization module brevitas.nn.QuantConv2d.
However, although this module itself allows for the specification of the total bit width, it does not allow for the explicit definition of the bit widths for the integer and fractional parts separately.
Therefore, by specifying the scaling type and other parameters as arguments, it is possible to perform fixed-point processing with any desired bit width.

\begin{figure}[t]
	\centering
	\includegraphics[width=\columnwidth]{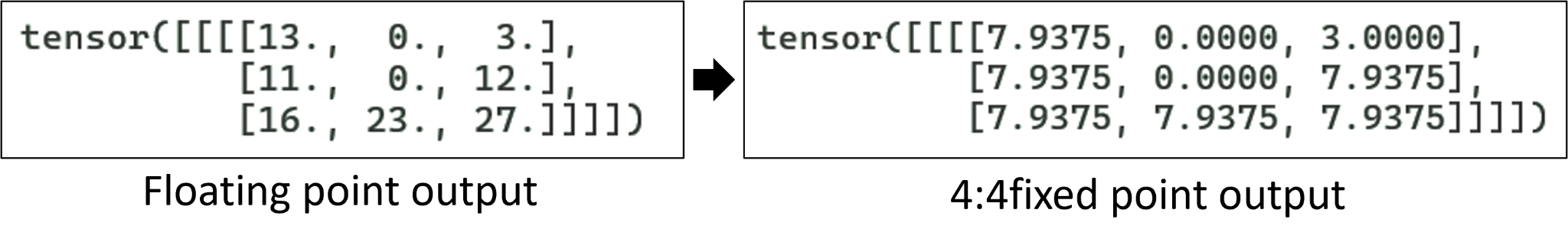}
	\caption{Comparison of outputs in the ReLU layer. Left: outputs with floating-point processing, Right: outputs with both integer and fractional parts quantized to 4 bits each.
			 In this case, the values can be represented in the range of -8.0 to 7.9375.}
	\label{output}
\end{figure}

Fig.\ref{output} shows the result of verifying whether fixed-point processing is performed correctly when the same implementation is applied to the ReLU process. 
The result confirms that the values are rounded to the range where they can be expressed according to the specified bit width.

\subsection{Post-Training Quantization}\label{PTQ_impl}
Unlike QAT, the PTQ processing flow is newly created, but some of the functions used in the flow are provided by PEFSL. The flow is as follows.
Here, the quantized model used for PTQ employs the same approach as QAT, where the convolution and activation layers of the ResNet model are replaced with Brevitas' quantization modules. 
Quantization is performed by applying this model to the pre-trained weights.

\begin{enumerate}
	\item Load the floating-point model (such as ResNet) and the weight data generated in advance through floating-point processing.
	\item Load the model implemented with fixed-point quantization.
	\item Transfer weights from the floating-point model to the fixed-point model.
	\item Use the fixed-point model to extract features from the training dataset and calculate their average vector.
	\item Use the fixed-point model to extract features from the test dataset and standardize the features using the average vector obtained in step 4.
	\item Measure accuracy through Nearest Class Mean.
\end{enumerate}

\subsection{Nearest Class Mean}
In the processing of NCM used as a classifier, the accuracy evaluation part is converted to fixed-point by quantizing the part that performs the computation of Equations (1) and (2).
Here, by using the brevitas.nn.QuantIdentity module, which directly quantizes the input and produces the output, we ensure that the results are equivalent to those obtained with fixed-point processing for each operation.

\section{Evaluation}
The datasets used for the evaluation were miniImageNet and CIFAR-FS, and the CNN model is ResNet12.
Table \ref{tab:mini_1-shot}-\ref{tab:cifar_5-shot} show the accuracy for floating-point processing and fixed-point processing (QAT and PTQ) with specified bit-widths for these datasets. 
The evaluations are based on few-shot learning, with the 1-shot case shown in Table \ref{tab:mini_1-shot}-\ref{tab:cifar_1-shot} and the 5-shot case in Table \ref{tab:mini_5-shot}-\ref{tab:cifar_5-shot}.
The data type for floating-point processing is torch.float32.

In PTQ, with a low bit width such as 3 bits for both integer and fractional parts, the accuracy drops by nearly 40\% compared to floating-point processing, while in QAT, the reduction is only about 6\%, reflecting the feature of QAT, which processes while considering the effects of quantization.
On the other hand, when the bit width is large, the effect of quantization becomes smaller, so the difference in accuracy between QAT and PTQ disappears.
From these results, it can be inferred that the fixed-point quantization implemented by both methods works well in this implementation.

With Tensil, which is used in the current design flow, the bit widths of the integer and fractional parts need to be 8 bits each or 16 bits each when implemented in hardware.
However, as seen from Table \ref{tab:mini_1-shot}, the accuracy comparable to floating point is almost maintained with only 5 bits each for QAT and 6 bits each for PTQ, with similar outcomes observed in the other tables.
These results indicate the possibility of further reduction of computational resources in hardware implementation.

\section{Conclusion}
In this paper, by implementing fixed-point processing across the entire design pipeline, we have ensured consistency in accuracy and further contributed to the creation of a versatile design and evaluation environment for edge AI hardware for few-shot learning.

However, the hardware implementation using Tensil faces several constraints. 
Therefore, we plan to further develop an environment that supports more flexible hardware design by utilizing alternative frameworks with greater adaptability in the future.

\begin{table}[t] 
	\centering
	\caption{Accuracy of QAT and PTQ on the miniImageNet Dataset (1-shot)}
	\label{tab:mini_1-shot}	
	\begin{tabular}{cccc}
	\toprule
	\multicolumn{2}{c}{Fixed Point} & \multirow{2}*{QAT} & \multirow{2}*{PTQ} \\
	int bit-width & frac bit-width &  & \\
	\midrule
	3 & 3 & $\textbf{54.54}\pm\textbf{0.15}$ & $\textbf{22.57}\pm\textbf{0.08}$ \\
	4 & 4 & $58.69\pm0.30$ & $46.30\pm0.19$ \\
	5 & 5 & $60.01\pm0.17$ & $57.06\pm0.19$ \\
	6 & 6 & $60.34\pm0.12$ & $59.00\pm0.20$  \\
	7 & 7 & $60.54\pm0.18$ & $59.52\pm0.20$ \\
	8 & 8 & $60.37\pm0.60$ & $59.70\pm0.20$  \\
	16 & 16 & $60.82\pm0.41$ & $59.80\pm0.20$  \\
	\midrule
	\multicolumn{2}{c}{Floating Point (32bit)} & \multicolumn{2}{c}{$60.35\pm0.17$} \\
	\bottomrule
	\end{tabular}
\end{table}

\begin{table}[t] 
	\centering
	\caption{Accuracy of QAT and PTQ on the CIFAR-FS Dataset (1-shot)}
	\label{tab:cifar_1-shot}	
	\begin{tabular}{cccc}
	\toprule
	\multicolumn{2}{c}{Fixed Point} & \multirow{2}*{QAT} & \multirow{2}*{PTQ} \\
	int bit-width & frac bit-width &  & \\
	\midrule
	3 & 3 & $\textbf{60.56}\pm\textbf{0.32}$ & $\textbf{26.76}\pm\textbf{0.13}$ \\
	4 & 4 & $65.67\pm0.62$ & $51.19\pm0.21$ \\
	5 & 5 & $67.08\pm0.12$ & $63.06\pm0.22$ \\
	6 & 6 & $67.00\pm0.66$ & $66.77\pm0.22$  \\
	7 & 7 & $67.43\pm0.19$ & $66.44\pm0.22$ \\
	8 & 8 & $66.86\pm0.44$ & $66.95\pm0.22$  \\
	16 & 16 & $67.19\pm0.41$ & $66.96\pm0.22$  \\
	\midrule
	\multicolumn{2}{c}{Floating Point (32bit)} & \multicolumn{2}{c}{$67.30\pm0.21$} \\
	\bottomrule
	\end{tabular}
\end{table}

\begin{table}[t] 
	\centering
	\caption{Accuracy of QAT and PTQ on the miniImageNet Dataset (5-shot)}
	\label{tab:mini_5-shot}	
	\begin{tabular}{cccc}
	\toprule
	\multicolumn{2}{c}{Fixed Point} & \multirow{2}*{QAT} & \multirow{2}*{PTQ} \\
	int bit-width & frac bit-width &  & \\
	\midrule
	3 & 3 & $\textbf{67.03}\pm\textbf{0.38}$ & $\textbf{44.89}\pm\textbf{0.17}$ \\
	4 & 4 & $74.97\pm0.23$ & $64.25\pm0.17$ \\
	5 & 5 & $76.86\pm0.18$ & $75.07\pm0.15$ \\
	6 & 6 & $77.42\pm0.15$ & $76.43\pm0.15$  \\
	7 & 7 & $77.47\pm0.03$ & $76.67\pm0.15$ \\
	8 & 8 & $77.28\pm0.28$ & $76.87\pm0.15$  \\
	16 & 16 & $77.62\pm0.21$ & $77.06\pm0.15$  \\
	\midrule
	\multicolumn{2}{c}{Floating Point (32bit)} & \multicolumn{2}{c}{$77.60\pm0.33$} \\
	\bottomrule
	\end{tabular}
\end{table}

\begin{table}[t] 
	\centering
	\caption{Accuracy of QAT and PTQ on the CIFAR-FS Dataset (5-shot)}
	\label{tab:cifar_5-shot}	
	\begin{tabular}{cccc}
	\toprule
	\multicolumn{2}{c}{Fixed Point} & \multirow{2}*{QAT} & \multirow{2}*{PTQ} \\
	int bit-width & frac bit-width &  & \\
	\midrule
	3 & 3 & $\textbf{73.51}\pm\textbf{0.38}$ & $\textbf{43.04}\pm\textbf{0.18}$ \\
	4 & 4 & $80.70\pm0.21$ & $69.19\pm0.18$ \\
	5 & 5 & $82.27\pm0.20$ & $79.44\pm0.17$ \\
	6 & 6 & $82.29\pm0.21$ & $82.31\pm0.16$  \\
	7 & 7 & $82.23\pm0.21$ & $82.23\pm0.16$ \\
	8 & 8 & $82.39\pm0.37$ & $82.48\pm0.16$  \\
	16 & 16 & $82.66\pm0.31$ & $82.57\pm0.16$  \\
	\midrule
	\multicolumn{2}{c}{Floating Point (32bit)} & \multicolumn{2}{c}{$82.58\pm0.55$}  \\
	\bottomrule
	\end{tabular}
\end{table}

\bibliographystyle{IEEEtran}
\bibliography{mwscas2024}

\section*{Acknowledgment}
This work was supported in part by JST CREST Grant Number JPMJCR19K.

\end{document}